\documentclass[twocolumn,prb]{revtex4}
\usepackage{graphics,graphicx}
\usepackage[utf8]{inputenc}
\usepackage{amsmath}
\usepackage{braket}   
\DeclareMathOperator{\sinc}{sinc} 
\usepackage{amsfonts}
\usepackage{amssymb}
\begin{document}
\title{Various facets of magnetic charge correlation: Micromagnetic and distorted wave Born approximation simulations study}
\author{G. Yumnam$^{1}$}
\author{J. Guo$^{1}$}
\author{D. K.~Singh$^{1,*}$}

\affiliation{$^{1}$Department of Physics and Astronomy, University of Missouri, Columbia, MO 65211}
\affiliation{$^{*}$email: singhdk@missouri.edu}

\begin{abstract}
{The emergent concept of magnetic charge quasi-particle provides a new realm to study the evolution of magnetic properties in two-dimensional artificially frustrated magnets. We report on the exploration of magnetic phases due to various magnetic charge correlation using the complementary numerical techniques of micromagnetic and distorted wave Born approximation simulations in artificial permalloy honeycomb lattice. The honeycomb element length varies between 10 nm and 100 nm, while the width and thickness are kept within the single domain limit. In addition to the charge ordered loop state, we observe disordered charge arrangement, characterized by the random distribution of $\pm$Q charges, in single domain size honeycomb lattice. As the length of honeycomb element increases, low multiplicity magnetic charges tend to form contiguous bands in thinner lattice. Thin honeycomb lattice with 100 nm element length exhibits a perfect spin ice pattern, which remains unaffected to the modest increase in the width of element size. We simulate scattering profiles under the pretext of distorted wave Born approximation formalism for the micromagnetic phases. The results are expected to provide useful guidance in the experimental investigation of magnetic phases in artificial honeycomb magnet.}

\end{abstract}

\maketitle

\section{Introduction}

The honeycomb lattice structure has generated significant research interest in recent times- primarily motivated by the unusual electronic and magnetic properties, as found in Graphene and nanostructured magnetic honeycomb lattice, respectively.\cite{Stamps,Peter,Summers2} Artificial magnetic honeycomb lattice is a prominent research venue to explore novel magnetism in reduced dimensionality. Originally conceived to explore the magnetic analogue of ice-rule and associated Dirac's effective monopoles using standard experimental techniques,\cite{Sondhi} such as magnetic force microscopy and X-ray dichroism method, it has become a subject of extensive investigation to find new properties of geometrically frustrated magnets.\cite{Peter} In recent years, artificially created magnetic honeycomb lattice is demonstrated to exhibit a broad and tunable range of novel magnetic phenomena that are difficult to achieve in a naturally occurring magnet, such as the entropy-driven magnetic charge-ordered state due to the spin chirality.\cite{Stamps,Chern} At low enough temperature, the magnetic correlation develops into a long range ordered spin solid state density, which is manifested by the periodic arrangement of the vortex loops of opposite chirality across the lattice.\cite{Glavic,Sendetsky} One of the underlying assumptions in the theoretical analysis of magnetic properties in two-dimensional honeycomb lattice is based on the proposition that a magnetic moment can be considered as a pair of magnetic charges of opposite polarities, as if it is a `dumbbell', that interact via the magnetic Coulomb interaction.\cite{Sondhi,Bramwell,Shen} Consequently, honeycomb vertices are occupied by two-types of magnetic charges: $\pm$3 and $\pm$1 units that are associated to the peculiar moment configurations where magnetic moment, aligned along the length of the element due to the shape anisotropy, either point to or away from the vertex at the same time or, two of them point to (or away) from the vertex and one points away (or to) from the vertex, respectively.\cite{Summers} These moment arrangements are also called `all-in or all-out' and `two-in \& one-out' (or vice-versa) spin configurations. 

\begin{figure*}
\centering
\includegraphics[width=18. cm]{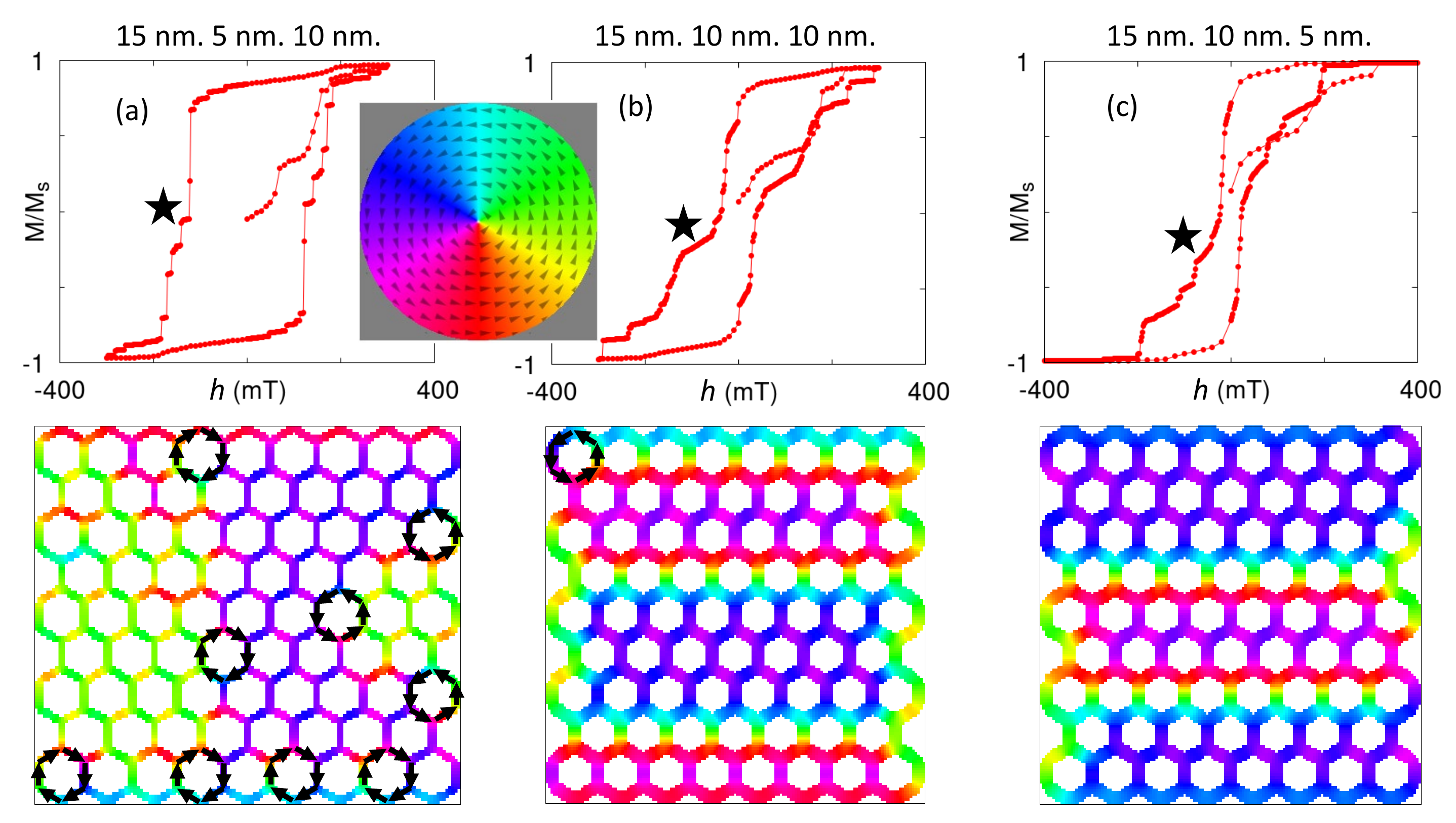} \vspace{-2mm}
\caption{Micromagnetic simulation of permalloy honeycomb lattice, made of single domain element size. Top panel- magnetic hysteresis loop generated by MM simulation for various element size honeycombs. Bottom panel - magnetic charge configuration after magnetic transition near zero magnetic field (marked by asterisk). (a) Element size of 15 nm $\times$ 5 nm $\times$ 10 nm. The system tends to develop the spin solid loop state in thin honeycomb lattice. Inset shows the color profile of magnetization direction. (b) Element size of 15 nm $\times$ 10 nm $\times$ 10 nm. As the lattice elements become broader, disordered configuration of magnetic charges arise. (c) Element sizes of 15 nm $\times$ 10 nm $\times$ 5 nm. Spin ice state is prominent in this case.} \vspace{-4mm}
\end{figure*}

\begin{figure*}
\centering
\includegraphics[width=18. cm]{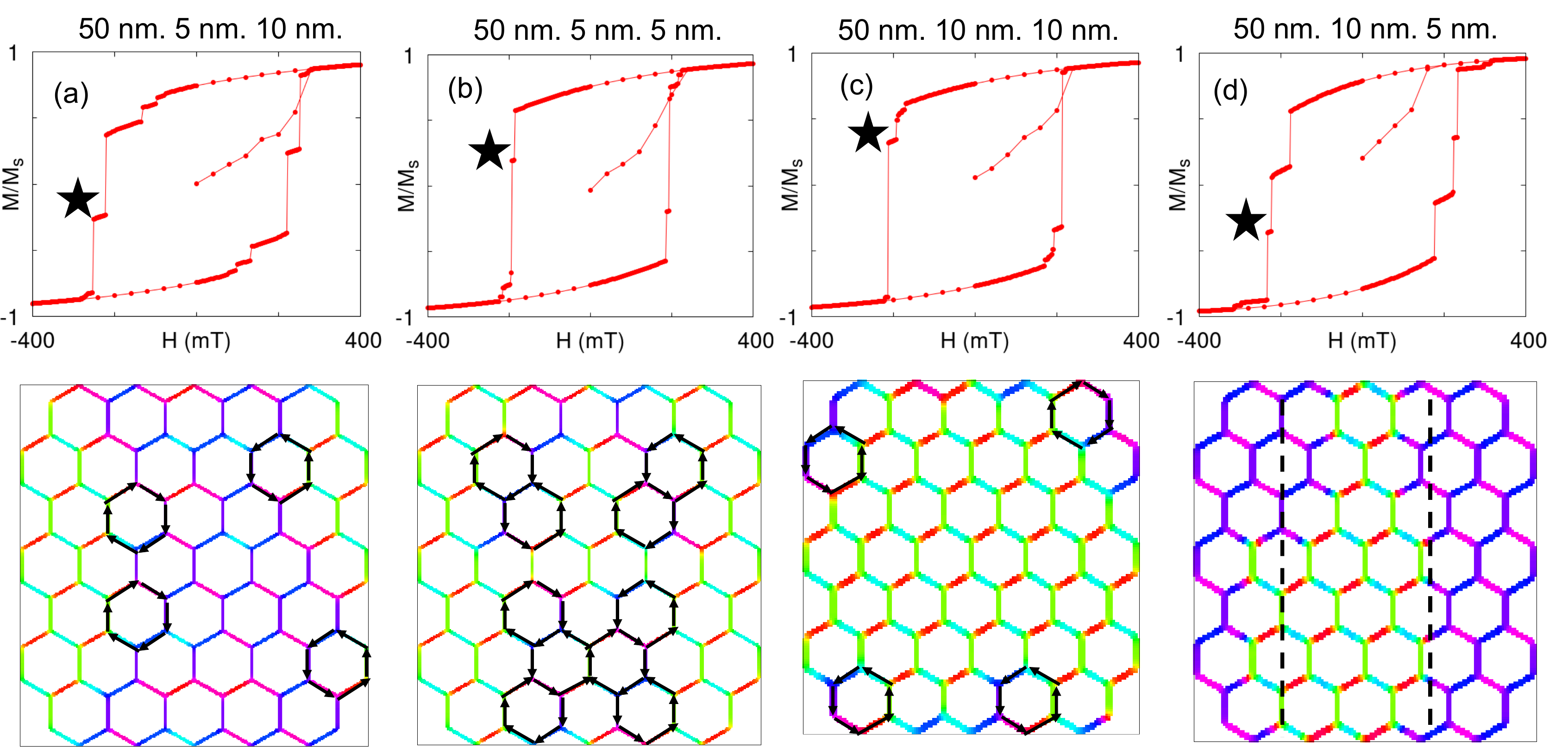} \vspace{-2mm}
\caption{MM simulation of permalloy honeycomb lattice with 50 nm element length. Top panel- magnetic hysteresis loop generated. Bottom panel - magnetic charge configuration after magnetic transition (marked by asterisk). (a) Element size of 50 nm $\times$ 5 nm $\times$ 10 nm. The system tends to develop the spin solid loop state as $T \rightarrow$ 0 K. (b) Element size of 50 nm $\times$ 5 nm $\times$ 5 nm. As the thickness of the honeycomb lattice decreases, more loops are formed and tend to be close to each other, the system tends to develop a complete spin solid loop state. (c) Element sizes of 50 nm $\times$ 10 nm $\times$ 10 nm. As the lattice  becomes wider, the system tends to develop spin ice state as $T \rightarrow$ 0 K. (d) Element sizes of 50 nm $\times$ 10 nm $\times$ 5 nm. Contiguous bands of $\pm$Q magnetic charges, reminiscent of stripe-type order, develop in thin honeycomb lattice with wider element.} \vspace{-4mm}
\end{figure*}

Magnetic charges are represented by the Pauli matrices quantum operator.\cite{Chern} The quantum mechanical properties of magnetic charge, also termed as quasi-particle,\cite{Stamps} enables the exploration of dynamic magnetic states in artificial Kagome ice.\cite{Rougemaille,Ferhan,De Long} One of such states is the quantum disordered magnetic ground state due to the competing energetics between the nearest neighbor and the next nearest neighbor exchange interactions ($J_1$ and $J_2$, respectively).\cite{Willis,Shastry} However, the thermal tunability of lattice magnetization is necessary to the realization of such novel state, as it facilitates magnetic charge dynamics to incite a massively degenerate ground state at $T \rightarrow$ 0 K.\cite{Chen2} Experimental evidence to this proposition was recently obtained in a magnetic honeycomb lattice with thermally tunable characteristic, made of single domain size connecting elements.\cite{Yumnam} Besides the emergent magnetic properties, magnetic charge quasi-particles are also found to develop the magnetic analogue of quintessential electronic state of Wigner crystal in the simultaneous applications of electric current and magnetic field.\cite{Chen} Magnetic charge's versatility in the manifestation of various ground state properties under different thermal, electrical and magnetic field tuning conditions have spurred a plethora of new researches. In this article, we report a systematic study of the evolution of magnetic charge correlation as a function of the geometrical tuning parameter e.g. variation in element size in artificial permalloy honeycomb lattice. The honeycomb element length varies between 10 nm to 100 nm, while the width and the thickness vary between the two limits of 5 nm and 10 nm. Since width and thickness dimensions are always smaller than the typical domain size in permalloy, $\sim$ 18 nm,\cite{Coey} the honeycomb elements in this study are either single domain or constricted single domain in at least two-directions. Micromagnetic simulations reveal that 5 nm thick honeycomb with 100 nm long elements develops a perfect spin ice state. As the element size decreases, the system exhibits a variety of interesting charge configurations that include the charge ordered loop state, extending to the spin solid state, disordered phase and a stripe-like contiguous bands of $\pm$Q charges in 50 nm element size lattice. The micromagnetic simulation study is complemented by the detailed modeling of scattering profiles using the distorted wave Born approximation (DWBA) formalism for the theoretically generated magnetic phases. The DWBA profiles can act as guide in the experimental investigation of theoretical phases using macroscopic probes, such as neutron reflectometry.

\begin{figure*}
\centering
\includegraphics[width=18. cm]{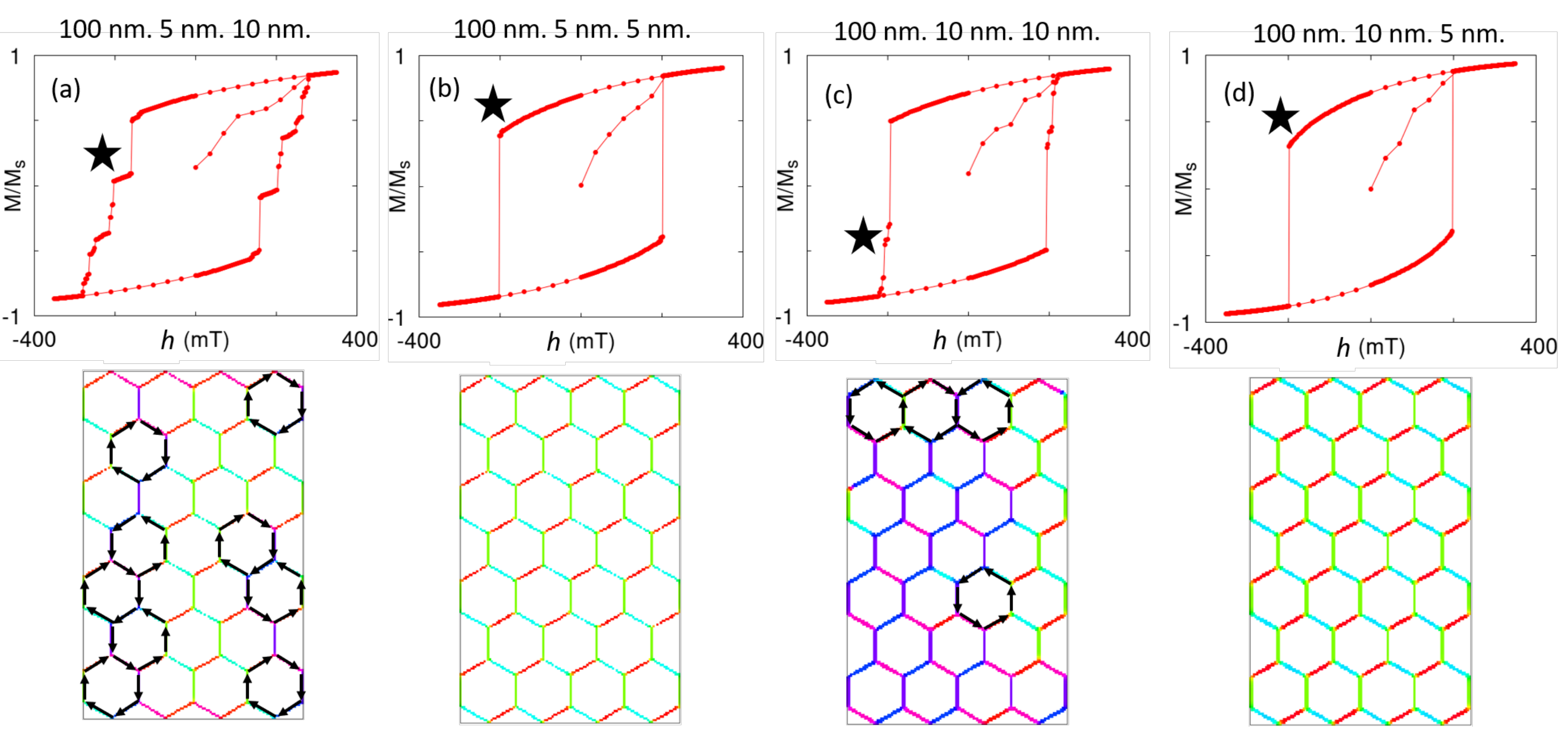} \vspace{-4mm}
\caption{MM simulation of 100 nm element length. (a) Element size of 100 nm $\times$ 5 nm $\times$ 10 nm. The system tends to develop the spin solid loop state as $T \rightarrow$ 0 K. (b) Element size of 100 nm $\times$ 5 nm $\times$ 5 nm. Magnetic phase near zero field is primarily dominated by the spin ice-type magnetic charge configuration. (c) Element size of 100 nm $\times$ 10 nm $\times$ 10 nm. As the lattice becomes wider, fewer loops are founded. (d) Element size of 100 nm $\times$ 10 nm $\times$ 5 nm. Near zero field, the magnetic phase in thin honeycomb lattice with wider elements is also primarily dominated by the spin ice-type magnetic charge configuration.} \vspace{-4mm}
\end{figure*}

There are several experimental methods to determine the nature of magnetic charge correlations or infer about the phase transition process in two-dimensional artificial spin ice. Some of the notable techniques include magnetic force microscopy, X-ray dichroism method, nonlinear susceptibility analysis, photoemission electron microscopy and polarized neutron reflectometry (PNR).\cite{Sendetsky,Peter,Glavic,Dahal,Ferhan} While some of these techniques are suitable for elucidating the local magnetic correlation, statistical probes, such as PNR method, are by design the bulk experimental procedure to deduce both the short range and the long range nature of correlations.\cite{Sinha} For instance, the spin solid state manifests the long range ordered arrangement of magnetic charges, Therefore, it is desirable to probe such characteristic in a sample consisting of the macroscopic ensemble of honeycomb lattice units using scattering method, such as PNR measurement technique. However, unlike magnetic force microscopy, neutron reflectometry measurement does not yield direct information regarding the magnetic correlation in real space. It is typically inferred from the modeling of the off-specular reflectometry profile using the established numerical approach of the distorted wave Born approximation (DWBA) formalism.\cite{Lauter,Glavic} Numerical modeling using the DWBA method is non-trivial. In this article, we present DWBA simulations of various charge correlated phases, as predicted by theoretical calculations\cite{Stamps} and micromagnetic simulations, in artificial magnetic (permalloy, Ni$_{0.8}$Fe$_{0.2}$) honeycomb lattice systems of varying element sizes. Here is the outline of the article: first, we describe the micromagnetic (MM) simulations in low temperature limit $T \rightarrow$ 0 K regarding the magnetic charge correlation on honeycomb vertices of varying element sizes. Second, DWBA simulated reflectometry profiles for various charge correlations are depicted using the contour maps in reciprocal space. For this purpose, some of the relevant parameters from the recently reported experimental results are utilized.\cite{Glavic,Summers,Yumnam} Finally, we summarize the results with a brief outlook for the future research.

\section{Micromagnetic Simulations}

Micromagnetic simulations are carried out using the Object Oriented MicroMagnetic Framework (OOMMF).\cite{OOMMF} Magnetic field dependent magnetization evolution, together with the flexible geometry specification, allows us to study the response of the magnetization as a function of magnetic field, applied in-plane to honeycomb thin film. The simulated geometry is made of honeycomb lattice with connected topography where the permalloy (Ni$_{0.8}$Fe$_{0.2}$) element length varies between 10 nm to 100 nm. For MM simulations, honeycomb lattices are discretized into grids with the individual mesh size of 2$\times$2$\times$1.25 nm$^{3}$ (X, Y and Z) and 2$\times$2$\times$2.5 nm$^{3}$ for thickness 5 nm and 10 nm, respectively. Nanostructured magnetic materials were previously simulated with similar grid and mesh sizes.\cite{Singh} The simulation utilizes the Landau-Lifshitz-Gilbert equation of magnetization relaxation in a damped medium. It is given by \cite{LaLiGi1,LaLiGi2,Brown},
\begin{eqnarray}
{\frac{\textbf{dm}}{dt}}&=& {-\gamma}{\textbf{m} \times \textbf{h}_{eff}}+ {\alpha}~ {\textbf{m}} \times \frac{\textbf{dm}}{dt},
\end{eqnarray}
where $\gamma$ is the gyromagnetic ratio and $\alpha$ is damping constant. The effective field is given by $\textbf{h}_{eff}(T=0)$ = $-\delta$$\textbf{H}$/$\delta m$. The Hamiltonian, $\textbf{H}$, of the system consists of four terms: exchange energy, uniaxial anisotropic energy, magnetostatic energy and the Zeeman energy. For MM simulations, we have used the typical values for permalloy material e.g. exchange stiffness $A$ = 1.0$\times$10$^{-11}$ J/m, saturation magnetization $M_s$ = 1.0$\times$10$^6$ A/m, uniaxial anisotropy strength $K_1$ = -5.0$\times$10$^{3}$ J/m$^{3}$, damping constant $\alpha$ = 0.2. During the simulation, the magnetic field is applied along the Y direction, and is incremented by varying step sizes in order to capture the transition states.   

\begin{figure*}
\centering
\includegraphics[width=18. cm]{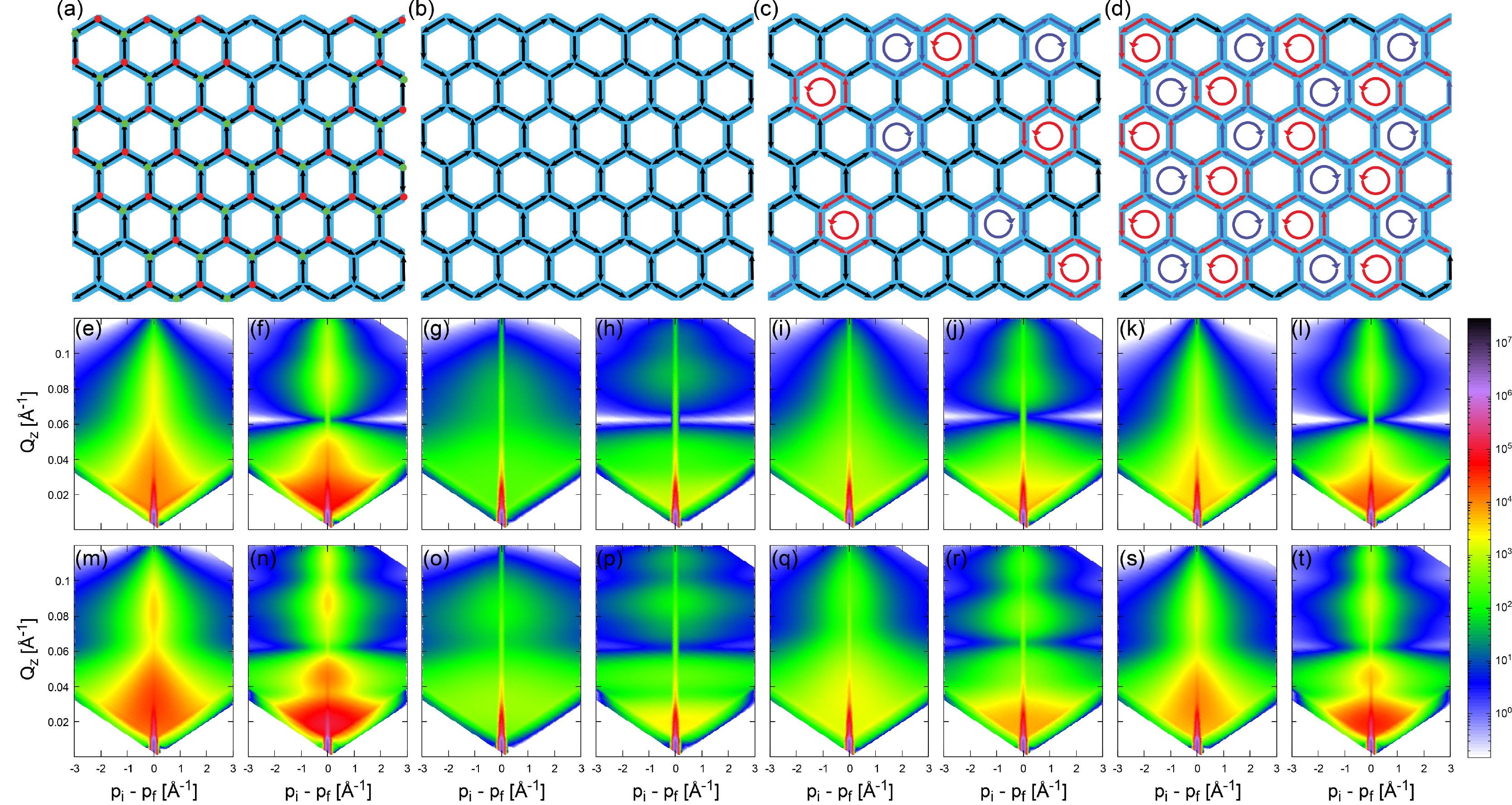} \vspace{-4mm}
\caption{DWBA  simulation of permalloy honeycomb lattice of 12 nm element length. (a-d) Top panel- magnetic charge configurations of disordered, spin ice, magnetic charge ordered and spin solid loop states. (e, f, m, n) DWBA simulation of disordered magnetic charge configuration (shown in fig. a) in various element size lattices of 12 nm $\times$ 5 nm $\times$ 5 nm, 12 nm $\times$ 5 nm $\times$ 10 nm, 12 nm $\times$ 10 nm $\times$ 5 nm, 12 nm $\times$ 10 nm $\times$ 10 nm, respectively. (g, h, o, p) DWBA simulation of spin ice configuration (shown in fig. b) in various element size lattices of 12 nm $\times$ 5 nm $\times$ 5 nm, 12 nm $\times$ 5 nm $\times$ 10 nm, 12 nm $\times$ 10 nm $\times$ 5 nm, 12 nm $\times$ 10 nm $\times$ 10 nm, respectively. (i, j, q, r) DWBA simulation of magnetic charge ordered state (shown in fig. c) in various element size lattices of 12 nm $\times$ 5 nm $\times$ 5 nm, 12 nm $\times$ 5 nm $\times$ 10 nm, 12 nm $\times$ 10 nm $\times$ 5 nm, 12 nm $\times$ 10 nm $\times$ 10 nm, respectively. (k, l, s, t) DWBA simulation of spin solid state (shown in fig. d) in various element size lattices of 12 nm $\times$ 5 nm $\times$ 5 nm, 12 nm $\times$ 5 nm $\times$ 10 nm, 12 nm $\times$ 10 nm $\times$ 5 nm, 12 nm $\times$ 10 nm $\times$ 10 nm, respectively} \vspace{-4mm}
\end{figure*}
 
The simulated hysteresis curves as a function of magnetic field and associated magnetic profiles for various element sizes are shown in Fig. 1-3. Qualitative differences between the magnetic hysteresis curves for various element size lattices are clearly noticeable. A multitude of magnetic phases tend to emerge near the zero field as the element size changes. The plot of M/M$_{s}$ vs $h$, M$_{s}$ and $h$ being saturation magnetization and magnetic field, respectively, depicts sharp transition near the zero field in single domain size element case where the geometrical parameters are smaller than the typical domain size in permalloy, $\sim$ 18 nm.\cite{Coey} The simulated magnetization profile in this state is characterized by the vortex configuration, which is the key element of the spin solid state. In this case , the system exhibits a tendency to attend the spin solid loop state or magnetic charge ordered state or a mixture of both. At moderate thickness and width, $t$ and $w$ = 10 nm, the density of magnetic vortex loop decreases. Basically, the magnetic charge profile manifests a disordered configuration. The simulated pattern of disordered state, consisting of only $\pm$Q charges, differs from a recent experimental report where both $\pm$Q and $\pm$3Q charges were found to randomly occupy the honeycomb vertices.\cite{Yumnam} The experimental observation was explained in terms of the competing energetics between the nearest neighbor and the next nearest neighbor exchange interactions. Micromagnetic simulations, on the other hand, does not take into account the next nearest neighbor exchange interaction. Perhaps, the inclusion of next nearest exchange interaction is more appropriate in this case, as the inter-elemental dipolar interaction energy is much smaller, $\sim$ 15 K. Hence, the next nearest exchange term can be a dominant term, compared to the dipolar term, in the Hamiltonian.

The development of charge ordered or loop state also seems to occur in honeycomb lattice with element size of 50 nm$\times$ 5 nm$\times$ 10 nm, see Fig. 2. The magnetic vortex loops become prevalent as the lattice becomes thinner, $t$ = 5 nm, tending to develop the spin solid state. Increasing the width and thickness of the honeycomb element seem to drive the vortex loops to the edges of the lattice. An interesting magnetic phase arises in honeycomb lattice with element size of 50 nm$\times$ 10 nm$\times$ 5 nm. MM simulations show that the contiguous bands of $\pm$Q charges develop. Clearly, this is not a disordered state. Rather, a peculiar type of low multiplicity charge ordering takes place. An entirely different scenario emerges in thin (5 nm) honeycomb lattice with long element e.g. 100 nm$\times$ 10 nm or 100 nm$\times$ 5 nm. Micromagnetic simulation suggests that the system exhibits a well-ordered pattern of $\pm$Q magnetic charges. As the lattice becomes thicker, the charge ordered vortex loops are formed across the plaquette. Perhaps, the simulation results can be different for a different set of exchange stiffness and uniaxial anisotropy strength parameters. However, for the same values of $A$ and $K_1$, it is inferred that the ratio of the width and the thickness of the honeycomb element plays crucial role in magnetic charge correlation on honeycomb vertices. 
 
Unlike the single domain element size lattice where the honeycomb element behaves as a single magnetic unit, relatively larger element length (50 nm or 100 nm) possibly sets the stage for the dominance of Bloch wall dynamics. So, even though the inter-elemental dipolar interaction energy is expected to be comparable in honeycomb lattice of fixed element length, say 50 nm, the remnant magnetization states are drastically different for different $w/t$ ratio. The direct experimental observation of the stripe bands or the disordered state of magnetic charges in honeycomb lattice with constricted single domain elements can be challenging. It is very difficult to resolve 5-10 nm structure using the MFM technique. Scattering method, such as polarized neutron reflectometry (PNR) or soft x-ray scattering, can be quite suitable for the experimental investigation of micromagnetic phases in these systems. In the case of PNR method, analysis of the off-specular data using the distorted wave Born approximation formalism can reveal the underlying magnetic charge configuration. In the following section, we show pertinent DWBA simulations of various magnetic charge states in artificial magnetic honeycomb lattice. The simulated results can be compared with experimental observations to deduce magnetic charge arrangement in magnetic honeycomb lattice.

\section{Distorted Wave Born Approximation Simulations}

\begin{figure*}
\centering
\includegraphics[width=18. cm]{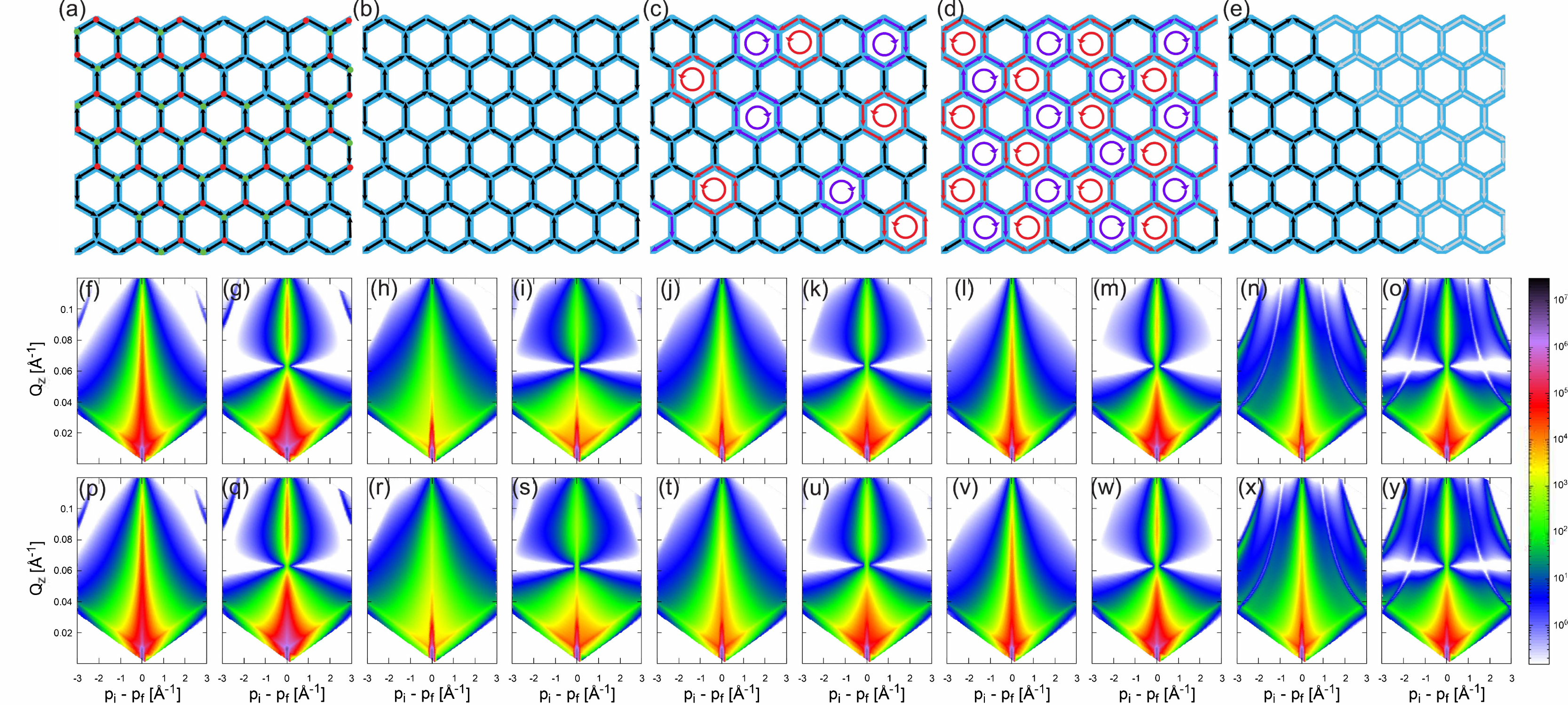} \vspace{-4mm}
\caption{DWBA  simulation of permalloy honeycomb lattice of 50 nm element length. (a-e) Top panel- magnetic charge configurations of disordered, spin ice, magnetic charge ordered, spin solid and low multiplicity magnetic charge stripe states. (f, g, p, q) DWBA simulation of disordered magnetic charge configuration (shown in fig. a) in various element size lattices of 50 nm $\times$ 5 nm $\times$ 5 nm, 50 nm $\times$ 5 nm $\times$ 10 nm, 50 nm $\times$ 10 nm $\times$ 5 nm, 50 nm $\times$ 10 nm $\times$ 10 nm, respectively. (h, i, r, s) Spin ice configuration (shown in fig. b) in various element size lattices of 50 nm $\times$ 5 nm $\times$ 5 nm, 50 nm $\times$ 5 nm $\times$ 10 nm, 50 nm $\times$ 10 nm $\times$ 5 nm, 50 nm $\times$ 10 nm $\times$ 10 nm, respectively. (j, k, t, u) Charge ordered configuration (shown in fig. c) in 50 nm $\times$ 5 nm $\times$ 5 nm, 50 nm $\times$ 5 nm $\times$ 10 nm, 50 nm $\times$ 10 nm $\times$ 5 nm, 50 nm $\times$ 10 nm $\times$ 10 nm, respectively. (l, m, v, w) Spin solid state (shown in fig. d) in various element size lattices of 50 nm $\times$ 5 nm $\times$ 5 nm, 50 nm $\times$ 5 nm $\times$ 10 nm, 50 nm $\times$ 10 nm $\times$ 5 nm, 50 nm $\times$ 10 nm $\times$ 10 nm, respectively. (n, o, x, y) Stripe state (shown in fig. e) in various element size lattices of 50 nm $\times$ 5 nm $\times$ 5 nm, 50 nm $\times$ 5 nm $\times$ 10 nm, 50 nm $\times$ 10 nm $\times$ 5 nm, 50 nm $\times$ 10 nm $\times$ 10 nm, respectively.} \vspace{-4mm}
\end{figure*}

\begin{figure*}
\centering
\includegraphics[width=18. cm]{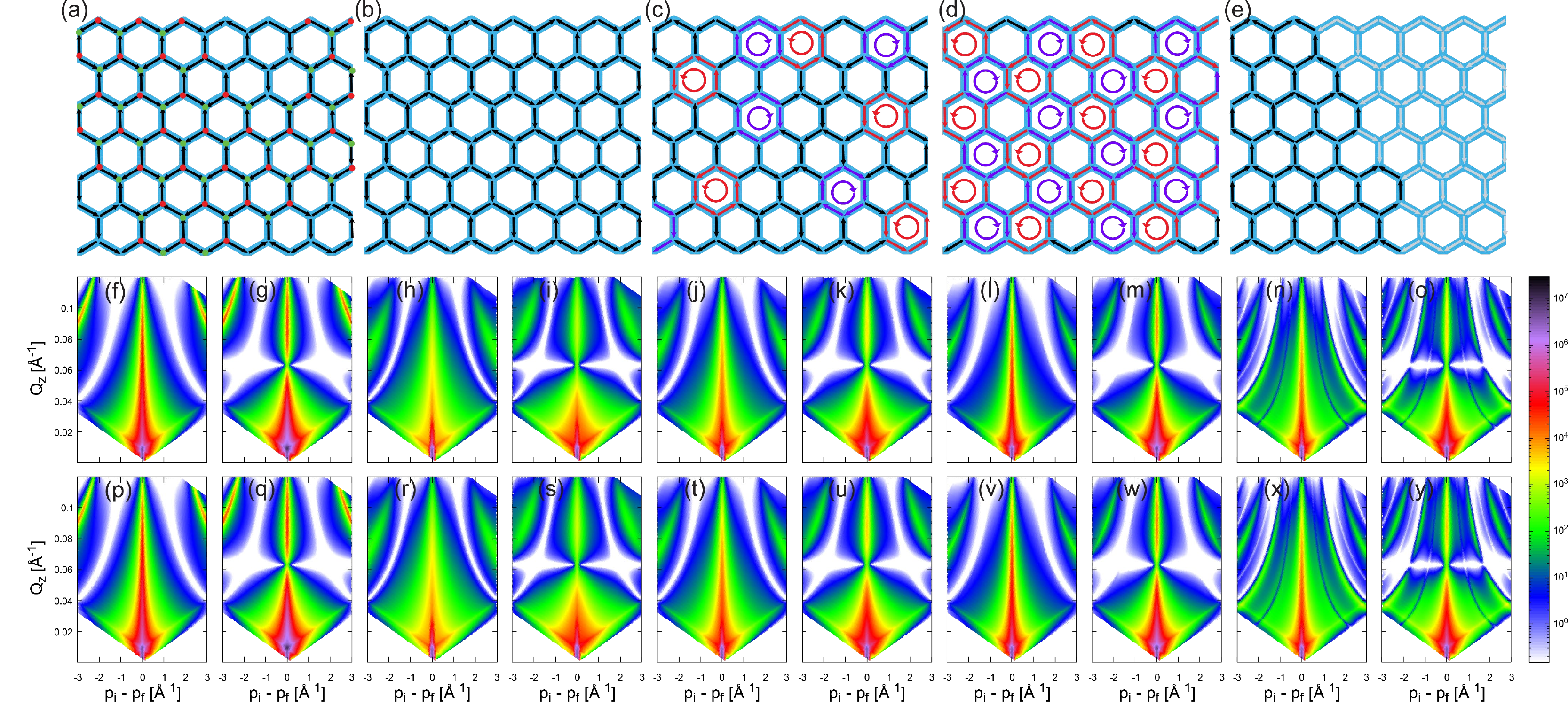} \vspace{-4mm}
\caption{DWBA  simulation of permalloy honeycomb lattice of 100 nm element length. (a-e) Top panel- magnetic charge configurations of disordered, spin ice, magnetic charge ordered, spin solid and low multiplicity magnetic charge stripe states. (f, g, p, q) DWBA simulation of disordered magnetic charge configuration (shown in fig. a) in various element size lattices of 100 nm $\times$ 5 nm $\times$ 5 nm, 100 nm $\times$ 5 nm $\times$ 10 nm, 100 nm $\times$ 10 nm $\times$ 5 nm, 100 nm $\times$ 10 nm $\times$ 10 nm, respectively. (h, i, r, s) Spin ice configuration (shown in fig. b) in various element size lattices of 100 nm $\times$ 5 nm $\times$ 5 nm, 100 nm $\times$ 5 nm $\times$ 10 nm, 100 nm $\times$ 10 nm $\times$ 5 nm, 100 nm $\times$ 10 nm $\times$ 10 nm, respectively. (j, k, t, u) Charge ordered configuration (shown in fig. c) in 100 nm $\times$ 5 nm $\times$ 5 nm, 100 nm $\times$ 5 nm $\times$ 10 nm, 100 nm $\times$ 10 nm $\times$ 5 nm, 100 nm $\times$ 10 nm $\times$ 10 nm, respectively. (l, m, v, w) Spin solid state (shown in fig. d) in various element size lattices of 100 nm $\times$ 5 nm $\times$ 5 nm, 100 nm $\times$ 5 nm $\times$ 10 nm, 100 nm $\times$ 10 nm $\times$ 5 nm, 100 nm $\times$ 10 nm $\times$ 10 nm, respectively. (n, o, x, y) Stripe state (shown in fig. e) in various element size lattices of 100 nm $\times$ 5 nm $\times$ 5 nm, 100 nm $\times$ 5 nm $\times$ 10 nm, 100 nm $\times$ 10 nm $\times$ 5 nm, 100 nm $\times$ 10 nm $\times$ 10 nm, respectively.} \vspace{-4mm}
\end{figure*}

Distorted wave Born approximation method relies on the discretization of experimental geometry into the scattering matrix. Basically, we define the honeycomb sample as a multilayer specimen with two layers (\textbf{l}) of magnetic film and silicon substrate. The scattering matrix elements for such a system can be described by:

\begin{equation*}
\braket{ \psi_i | \delta v | \psi_f } = \sum_{\textbf{l}} \sum_{\pm i} \sum_{\pm f} \braket{ \psi^\pm_{i\textbf{l}} |\delta v | \psi^\pm_{f\textbf{l}}}
\end{equation*}
where, $\delta v$ is the first order perturbation expansion term of scattering length density $\left(v(\mathbf{r})\right)$, and $\psi_i$, $\psi_f$ denotes the incident, and final wave functions, respectively. The forward or backward traveling wave function in real-space is given by $\psi^+$  and $\psi^-$, respectively. While the bottom-layer consists of nanostructured silicon with honeycomb pattern, the top layer is made of permalloy honeycomb lattice. We introduce a layout of honeycomb pattern of permalloy-hexagons with cylinders cut-out from the center with a lattice spacing of $a=31$ nm within the permalloy-layer. The form factor for the cylindrical unit is defined as $F = 2\pi R^2 t \sinc \left( \dfrac{q_zt}{2} \right)\exp\left(\dfrac{iq_zt}{2} \right) \dfrac{J_1(q_{||}R)}{q_{||}R}$, where, $q_{||} \equiv \sqrt{q_x^2 + q_y^2}$ and $J_1$ is a Bessel function of the first-kind. The other parameters are radius $R$ and height $t$ that depend on the size of the honeycomb element. The magnetic phases, as inferred from the micromagnetic simulations, were constructed by using the rectangular elements with fixed magnetization directed along its length. The form factor of the rectangular element is defined by, 
$F = lwt \sinc\left(\dfrac{q_xl}{2}\right)\sinc\left(\dfrac{q_yw}{2}\right)\sinc\left(\dfrac{q_zt}{2}\right)\exp\left(\dfrac{iq_zt}{2}\right)$, where $l$, $w$, and $t$ denote length, width and height, respectively. The magnetization elements in the hexagonal lattice develop long-range correlations via the inter-cluster interference. Correspondingly, the scattering matrix element can be written as,
\begin{eqnarray*}
\braket{ \psi_i | \delta v | \psi_f } = \sum_j \exp\left({ i\mathbf{q}_{||}\mathbf{R}_{j||}}\right) \\
        \int d^2r_{||}\exp\left({i\mathbf{q}_{||}\mathbf{r}_{||}}\right) \int dz \phi^*_i(z) F(\mathbf{r} -\mathbf{R}_{j||}; \mathbf{T}_j)\phi_f(z) 
\end{eqnarray*}
where, $F(\mathbf{r}-\mathbf{R}_{ij}; \mathbf{T}_{j})$ is the form-factor for the $\textbf{j}^{\text{th}}$ element, such that $v_p(\mathbf{r}) = \sum_j F(\mathbf{r}-\mathbf{R}_{j||}; \mathbf{T}_j)$. The elastic scattering cross-section is given by, $\dfrac{d\sigma}{d\Omega} = |\braket{\psi_i | \delta v | \psi_f}|^2$. To account for the finite-size effect, we have used a 2D lattice interference function with a large isotropic 2D-Cauchy decay function with the lateral structural correlation lengths of $\lambda_{x,y} = 1$, $5$ and $10$ $\mu$m for the case of $l$ = $12$, $50$ and $100$ nm, respectively. The position-correlation is given as $\rho_S G(\mathbf{r}) = \sum_{m,n} \delta\left( \mathbf{r} - m\mathbf{a} -n\mathbf{b} \right) - \delta\left(\mathbf{r}\right)$, with lattice basis (\textbf{a, b}). We have also introduced the effects of natural-disorder in the system by applying a small Debye-Waller factor corresponding to a position-variance of $\braket{\mathbf{x}}^2 = 1 $ nm$^2$. The interference function can be written as: 
\begin{equation*}
S({q}) = \rho_S \sum_{q_i \in \Lambda^*} \dfrac{2\pi\lambda_x\lambda_y}{\left(1+q_x^2\lambda_x^2 + q_y^2\lambda^2_y\right)^{{3}/{2}}}
\end{equation*}

The simulated off-specular reflectivity profile is generated by using the DWBA modeling, implemented in the BornAgain~\cite{bornagain} software.\cite{Glavic} In Fig. 4-6, we show the simulated plots of off-specular reflectivity for various magnetic charge states of spin ice, magnetic charge ordered state, spin solid loop state, disordered state and the stripe phase in permalloy honeycomb lattice. The spin solid state is modeled by arranging the vortex loops of opposite chirality in an alternating order. In all plots, the y-axis represents the out-of-plane scattering vector ($Q_z$= $\frac{2\pi}{\lambda} (\sin{\alpha_i} + \sin{\alpha_f})$) whereas the difference between the z-components of the incident and the outgoing wave vectors ($p_i - p_f$ = $\frac{2\pi}{\lambda}(\sin{\alpha_i} - \sin{\alpha_f})$) is drawn along the x-axis. Thus, vertical and horizontal directions correspond to the out-of-plane and in-plane correlations, respectively.\cite{Lauter}. The specular reflectivity lies along the x = 0 line.

As shown in the lower panel of Fig. 4, significant off-specular scattering develops due to the spin-spin correlation in honeycomb lattice of 12 nm element length. The simulated patterns exhibit broad but distinct bands of diffuse scattering along the x-axis. In the theoretically predicted spin solid state, the off-specular diffuse scattering is prevalent between Q$_{z}$ = 0.06 to 1 $\AA^{-1}$. Similar behavior was reported in the PNR measurements on artificial permalloy honeycomb lattice of similar ultra-small element size ($\sim$ 12 nm in length).\cite{Glavic} Unlike the spin solid or the magnetic charge ordered states, the off-specular reflection is significantly broader in the case of the disordered phase, consisting of both $\pm$Q and $\pm$3Q magnetic charges. Additionally, the diffuse scattering intensity around the specular line tends to develop the localized pattern along the Z-axis with the increasing thickness of the lattice, indicating the onset of finite size correlation. The simulation also reveals the constricted nature of diffuse scattering in both the magnetic charge ordered and the spin solid states in thicker lattice with 10 nm element length, compared to the thinner lattice. The 10 nm thick lattice with wider elements, $w$ = 10 nm, exhibits Q$_{z}$-dependence of diffuse scattering.

DWBA simulation results of 50 nm element size honeycomb lattice, along with the magnetic charge configurations, are shown in Fig. 5. Unlike in the case of 10 nm element size honeycomb, the diffuse scattering tends to shrink along the Q$_{z}$ direction in this case. Also, the qualitative difference between the reflectometry profiles for the spin ice and the spin solid phases becomes weaker as the lattice becomes thinner (5 nm thickness). In 10 nm thick lattice, there is an observable distinction between the three magnetic phases of spin ice, charge ordered state and the spin solid state. However, once again, the disordered phase manifests stronger diffuse scattering, compared to the theoretically predicted states. Micromagnetic simulations revealed that 50 nm element size honeycomb lattice tends to develop contiguous bands of $\pm$Q magnetic charges, resembling a stripe-like pattern, in thinner lattice. Simulated reflectometry profiles for this state for different lattice thicknesses are shown in Fig. 5e and 5i . In this case, we observe the patches of distinct diffuse scattering along the x-axis, indicating q-dependent in-plane correlation. As honeycomb element becomes wider, $w$ = 10 nm, the diffuse scattering assumes more conical shape. Finally, we show the DWBA simulated plots of 100 nm element length honeycomb lattice in Fig. 6. The width and thickness of honeycomb element are kept in the single domain limit. Therefore, a honeycomb element is multi-domain along the length, but exhibits the single domain characteristic along width and thickness directions. We observe strong qualitative difference between the simulated profiles for thin (5 nm) and thick (10 nm) lattices. However, the distinction between the theoretically predicted phases of spin ice and spin solid or the disordered state is apparently very weak for a given thickness of the lattice. Also, numerical results do not seem to be much affected by a variation in width of the honeycomb element (from 5 nm to 10 nm). Perhaps, the simulated reflectometry profiles can exhibit distinct patterns for larger thicknesses or broader elements of the lattice. We have not explored the multi-domain structure of honeycomb elements in all three directions. In the latter case, domain wall motion along different directions makes the micromagnetic simulation analysis more cumbersome. 

\section{Discussion}

Artificial magnetic honeycomb lattice is known to exhibit a plethora of theoretically predicted emergent phases as functions of temperature and magnetic field.\cite{Stamps} Micromagnetic simulations on honeycomb lattice of relatively smaller element sizes reveal additional magnetic phases of disordered charge configuration and contiguous bands of $\pm$Q charges (termed as stripe-type phase) in the remnant state. While the disordered state consists of a random distribution of $\pm$Q and $\pm$3Q charges, the stripe phase is characterized by the coexisting bands of ordered patterns of $+$Q and $-$Q charges. Experimental confirmation to a disordered ground state was recently reported in permalloy honeycomb lattice of single domain size element.\cite{Yumnam} Magnetic charges manifest highly dynamic behavior to the lowest measurement temperature in the disordered phase.\cite{Chen2} Experimental observation of the stripe-type phase of low multiplicity magnetic charges can be very exciting. 

Experimental investigation of magnetic states in artificial honeycomb lattice, made of single domain size magnetic elements, is a challenging task. Unlike the case of large element size magnetic honeycomb where the magnetic force microscopy (MFM) has proved to be an effective probe, the direct imaging of magnetic charges in such small geometry is not feasible. The spatial resolution ($\sim$ 50 nm) of MFM is larger than the geometrical dimension of individual element in the single domain limit. The scattering method, such as PNR, provides an appropriate platform to the experimental quest. Neutron reflectometry method, as available on MagRef instrument on BL-4A beam line at the Spallation Neutron Source at the Oak Ridge National Laboratory, is a versatile probe to study magnetic charge correlation in artificial magnetic lattice of single domain size elements. While the specular reflection provides information about the underlying magnetism and the magnetic layer thickness, the off-specular data can be used to infer the nature of planar correlation of magnetic charges in the lattice. However, modeling of the off-specular data is not trivial. DWBA method is a commonly used numerical technique to extract the intended information. The simulation results, presented in this article, are expected to provide the useful guide in this pursuit. Further theoretical researches on understanding the development of stripe charge ordered phase in artificial magnetic honeycomb lattice are highly desirable.

Research at the University of Missouri is supported by the U.S. Department of Energy, Office of Basic Energy Sciences under Grant No. DE-SC0014461.

\clearpage

\end{document}